\documentstyle[12pt,epsf,rotate]{article}

\newcommand{\n}{\hspace*{-2.5mm}}

\def\PR  #1 #2 #3 {Phys.\ Rev.~{\bf #1}, #2 (#3)}
\def\PRL #1 #2 #3 {Phys.\ Rev.~Lett.~{\bf #1}, #2 (#3)}
\def\PRD #1 #2 #3 {Phys.\ Rev.~D~{\bf #1}, #2 (#3)}
\def\PLB #1 #2 #3 {Phys.\ Lett.~{\bf B#1}, #2 (#3)}
\def\NPB #1 #2 #3 {Nucl.\ Phys.~{\bf B#1}, #2 (#3)}
\def\RMP #1 #2 #3 {Rev.~Mod.~Phys.~{\bf #1}, #2 (#3)}
\def\ZPC #1 #2 #3 {Z.~Phys.~C~{\bf #1}, #2 (#3)}

\begin{document}

\title{
\vskip-3cm{\baselineskip14pt
\centerline{\normalsize\hfill TUM--HEP--252/96}
\centerline{\normalsize\hfill September 1996}}
\vskip1.5cm
Perturbation Theory and Its Limitations\\ 
in the Higgs Sector of the SM
\thanks{Lecture given at the
{\it XXXVI Cracow School of Theoretical Physics}, Zakopane, Poland,
June 1--10, 1996.}
}

\author{
{\sc Kurt Riesselmann}\\
{\normalsize Institut f\"ur Theoretische Physik,
Technische Universit\"at M\"unchen,}\\
{\normalsize James-Franck Strasse, 85747 Garching, Germany}
}

\date{}

\maketitle

\begin{abstract}
This lecture reviews various Higgs-sector amplitudes which have been
calculated to two loops in the Higgs quartic coupling.  After
explaining the framework of these calculations, the perturbative
behaviour of the amplitudes is discussed, and perturbative upper
bounds on the Higgs boson mass are given.
\end{abstract}

\section{Introduction}

The simplest model of breaking the electroweak gauge symmetry ${\rm
SU}(2)_L\times {\rm U}(1)_Y$ spontaneously is the standard Higgs
model.  It consists of the spin-zero Higgs boson and three massless
Goldstone bosons, the latter ultimately being absorbed by the weak
gauge bosons. The standard Higgs model is regarded to be an effective
theory, only valid up to a cutoff energy $\Lambda$.  The maximal
allowed value of $\Lambda$ depends on the Higgs value of the Higgs
mass $M_H$, and it is connected to the renormalization group (RG)
behaviour of the Higgs sector.

In the absence of a more complete theory, it is important to
understand the perturbative limitations of the standard Higgs model:
if $M_H$ is too large, perturbation theory ceases to be a useful tool
for calculating physical observables of the Higgs sector, such as
cross sections and Higgs decay width.  Here we review the derivation
of upper perturbative bound on $M_H$.

In Sect.~\ref{basics} we briefly present the framework of the Higgs
sector, introducing the Lagrangian and the Higgs running
coupling. Sect.~\ref{pertlim} lists the various criteria which can be
used to judge breakdown of perturbation theory, and the results for
perturbative $M_H$ upper bounds are derived. A short discussion and
comparison with lattice results are given in Sect.~\ref{disc}.

\section{Basics}
\label{basics}

The Lagrangian of the standard Higgs sector is
\begin{equation}
{\cal L}_H =
{\textstyle\frac{1}{2}}\left(\partial_\mu\Phi\right)^\dagger
\left(\partial^\mu\Phi\right)-{\textstyle\frac{1}{4}}\lambda
\left(\Phi^\dagger\Phi\right)^2+
{\textstyle\frac{1}{2}}\mu^2\Phi^\dagger\Phi\,, \label{l0higgs}
\end{equation}
where
\begin{equation}
\Phi = \left(
\begin{array}{c}
        w_1+iw_2\\
        h+iz
\end{array}
\right)=\left(
\begin{array}{c}
        \sqrt{2}w^{+}\\
        h+iz
\end{array}
\right). \label{phi} 
\end{equation}
The doublet $\Phi$ has a nonzero expectation
value in the physical vacuum,
\begin{equation}
\langle\,\Omega\,|\,\Phi^\dagger\Phi\,|\,\Omega\,\rangle=v^2\,. \label{vev}
\end{equation}
To facilitate perturbative calculations, the field $h$ is expanded
around the physical vacuum, absorbing the vacuum expectation value by
the shift $h\rightarrow H+v$. Hence the field $H$ has zero vacuum
expectation value.  Rewriting Eq.~(\ref{l0higgs}), the Higgs Lagrangian 
has the form
\begin{eqnarray}
{\cal L}_H &=& 
{\textstyle\frac{1}{2}}\partial_\mu{\bf w}\cdot\partial^\mu{\bf w}
+{\textstyle\frac{1}{2}}\partial_\mu H\,\partial^\mu H
-{\textstyle\frac{1}{2}}M_H^2 H^2 + {\cal L}_{3pt} +{\cal L}_{4pt}\,, 
\label{lagrhiggs}
\end{eqnarray}
with the three-point and four-point interactions of the fields given by
\begin{eqnarray}
{\cal L}_{3pt}&=&-\lambda v\left({\bf w}^2H + H^3\right)\,,\\
{\cal L}_{4pt}&=&
-{\textstyle\frac{1}{4}}\lambda\left({\bf w}^4+2{\bf w}^2H^2+H^4
\right) \,.
\label{l1higgs}
\end{eqnarray}
Here $\bf w$ is the SO(3) vector of Goldstone scalars, $(w_1,\,w_2,\,w_3)$, 
with $w_3=z$. The
tadpole term and an additive constant have been dropped. Looking at
Eq.~(\ref{lagrhiggs}), the $w^\pm$ and $z$ bosons are massless, in
agreement with the Goldstone theorem \cite{goldstone}. The Higgs mass
$M_H$ and the Higgs quartic coupling $\lambda$ are related by
\begin{equation}
\lambda=M_H^2/2v^2=G_FM_H^2/\sqrt{2},\label{lambda}
\end{equation} 
where $G_F$ is the Fermi constant, and $v$ is the physical vacuum
expectation value, $v=2^{-1/4}G_F^{-1/2}=246$ GeV.  The Lagrangian
${\cal L}_H$ corresponds to taking the limit of vanishing gauge and
Yukawa couplings of the full standard model Lagrangian, and it is the
starting point for carrying out calculations using the equivalence
theorem \cite{lee,eqt1}.  The
equivalence theorem can also be used to calculate radiative corrections
without having to use the full SM Lagrangian \cite{eqt2}, but the use
of proper renormalization conditions is crucial \cite{BS}.

The standard Higgs model Lagrangian contains only one degree of
freedom. Once $M_H$ is determined experimentally, the coupling
$\lambda$ is fixed according to Eq.~(\ref{lambda}).  This will be one
of the many predictions to be tested experimentally when a Higgs
particle has been discovered.

Calculating physical observables with energy scales larger than $M_H$,
renormalization group methods suggest the use of the running coupling
$\lambda(\mu)$.  The evolution of $\lambda(\mu)$ as a function of
$\mu$ is given by the renormalization group equation
\begin{equation}
\label{rge}
\frac{d\lambda(\mu)}{d\ln\mu} = \beta_\lambda(\lambda(\mu),g_t(\mu),\dots)\;,
\end{equation}
with the initial condition at scale $\mu=M_H$ imposed by Eq.~(\ref{lambda}).  
At one loop, the beta function is
\begin{equation}
\beta_\lambda=24\lambda^2+12\lambda g_t^2 - 6g_t^4 + ...,
\end{equation}
and we find $\beta_\lambda>0$ for $M_H> 208$ GeV, assuming $m_t=175$
GeV.  For even larger values of $M_H$, the beta function is dominated
by the contributions from the Higgs coupling $\lambda\propto M_H^2$,
and we can neglect the top Yukawa coupling, $g_t$, as well as other
contributions.  In this limit, the three-loop beta function for the
Higgs coupling is \cite{LW2,NR}:
\begin{equation}
\beta(\lambda)
=24\frac{\lambda^2}{16\pi^2} 
\left[ 1 - 13\frac{\lambda}{16\pi^2}
         + 176.6\left(\frac{\lambda}{16\pi^2}\right)^2
\right]\;.
\label{beta}
\end{equation}
Neglecting the appropriate powers of $\lambda$, the above equations
determine the $n$-loop running coupling for $n\leq3$. Explicitly, the
one-loop heavy-Higgs running coupling is
\begin{eqnarray}
\lambda(\mu)&=&\frac{\lambda(M_H)}
{1 - 12\,\frac{\lambda(M_H)}{16\pi^2}\ln\left(\frac{\mu^2}{M_H^2}\right) }\,.
\label{run1lp}
\end{eqnarray}

\section{Limitations of perturbation theory}
\label{pertlim}

There are two scenarios in which perturbation theory can break down.  First,
the Higgs coupling can be nonperturbative for large values of $M_H$; see
Eq.~(\ref{lambda}). Secondly, the Higgs {\it running} coupling can be
nonperturbative if $\beta_\lambda>0$ and $\mu$ is large.  Increasing the scale
$\mu$ in Eq.~(\ref{run1lp}), $\lambda(\mu)$ eventually becomes infinite,
indicating the one-loop Landau pole.\footnote{ In the case of
  $\beta_\lambda<0$, increasing $\mu$ eventually leads to a negative Higgs
  running coupling.  This observation is related to the problem of vacuum
  stability. See \cite{sher} for a review.} Perturbation theory ceases to be
reliable long before reaching the location of this pole.

To judge for which values of the (running) Higgs coupling the
interactions become strong, one calculates perturbative amplitudes,
assuming small perturbative couplings.  Increasing $\lambda\propto
M_H^2$, the perturbative expansion will eventually break down, giving
upper perturbative bounds on the (running) Higgs coupling.

One has to distinguish two different kinds of physical observables.
Low-energy quantities such as Higgs decay amplitudes do not require RG
methods, hence no running coupling is involved. Therefore they
immediately yield an upper perturbative bound on $M_H$ via
Eq.~(\ref{lambda}).  On the other hand, cross sections with large
center-of-mass energy, $\sqrt{s}\gg M_H$, do require the use of the
running coupling.  Hence the onset of nonperturbative effects in such
processes is a function of both $M_H$ and $\mu=O(\sqrt{s})$.

There are several criteria we can 
use to judge the convergence of perturbation theory:
\begin{itemize}
\item The size of radiative corrections should be small,
   with higher order contributions being suppressed.
\item The renormalization scheme dependence ({\it e.g.,}
   $\overline{\rm MS}$ {\it vs.} OMS) of physical amplitudes should diminish
   when including higher order corrections.
\item The dependence of an amplitude on the scale $\mu$ should decrease
   with increasing order in perturbation theory.
\item Scattering amplitudes should not violate perturbative unitarity
   by a large amount.
\end{itemize}

{\it Exact} amplitudes, of course, have {\it no} scheme dependence,
{\it no} scale dependence, and feature {\it no} unitarity violations.
Converging perturbative amplitudes should therefore approach
this limit reasonably fast. Yet there are also limitations to
nonperturbative calculations in the SM Higgs sector.  Since the heavy-Higgs
running coupling has a positive beta function even in the
nonperturbative approach \cite{LW2,LW1,GKNZ}, 
there remains a cutoff scale $\Lambda$
beyond which the theory fails.  In return, this cutoff scale bounds
the possible Higgs mass range from above. We will return to this aspect
in Sect.~\ref{disc}.

\subsection{Perturbative bounds from Higgs decays}

A heavy Higgs particle ($M_H>2m_t$) decays mostly into pairs of
(longitudinally polarized) gauge bosons.  The branching ratio for
decays into a pair of top quarks is about 10\%.  Neglecting the
subleading corrections due to gauge and Yukawa couplings, the two-loop
$O(\lambda^2)$ corrections to these decay channels have been
calculated.  Using the OMS renormalization scheme, the bosonic decay
channel receives corrections\cite{hww1,hww2}
\begin{eqnarray}
\label{hwwwidth}
\Gamma(H&\n\rightarrow\n& ZZ,\;W^+W^-) \propto 
\lambda(M_H)\left(1
+ 2.8 \frac{\lambda(M_H)}{16\pi^2} 
+ 62.1 \frac{\lambda^2(M_H)}{(16\pi^2)^2}\right)\,,
\end{eqnarray}
and the two-loop result for the fermionic decay width is \cite{hff1,hff2}
\begin{eqnarray}
\Gamma(H&\n\rightarrow\n& t\bar t) \propto
g_t^2  \left(1
+ 2.1\frac{\lambda(M_H)}{16\pi^2}
- 32.7 \frac{\lambda^2(M_H)}{(16\pi^2)^2}\right)\,.
\label{hffwidth}
\end{eqnarray}
Comparing the magnitude of the one-loop and two-loop corrections, perturbation
theory seems to work up to values of $\lambda(M_H)\approx 7$, that is, Higgs
masses of about 1 TeV.  However, conversion of the results into the
$\overline{\rm MS}$ scheme as well as an analysis of the scale dependence using
RG methods reveals that higher-order terms may spoil perturbation theory
already for values of $\lambda(M_H)\approx 4$, or equivalently, $M_H\approx
700$ GeV \cite{NR}.  The principles of such analyses will be discussed in the
following section in connection with cross sections rather than decay widths.



\subsection{Perturbative bounds from scattering processes}

\begin{figure}[tb]
\vspace*{13pt}
\centerline{
\epsfysize=2.5in \rotate[l]{\epsffile{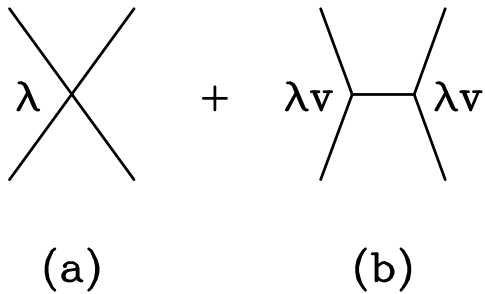}}
}
\vspace{0.2in}
\caption{ The two Feynman graphs contributing to the
tree-level $O(\lambda)$ term of the $W_L^+W_L^-\rightarrow Z_LZ_L$ amplitude.}
\label{feynman}
\end{figure}

A typical scattering process related to the SM Higgs sector is
$W^+W^-\rightarrow ZZ$.  This process is of interest for future
colliders such as the LHC or linear $e^+e^-$ and $\mu^+\mu^-$
colliders.  At tree level, the amplitude consists of a gauge
contribution $O(g^2)$ and a Higgs sector contribution $O(\lambda)$.
The $O(\lambda)$ term is connected to the longitudinally polarized
component of the gauge bosons. It can be calculated using ${\cal
L}_H$, receiving contributions from both three-point and four-point
interactions, see Fig.~\ref{feynman}.

\begin{figure}[bt]
\vspace*{13pt}
\centerline{
\epsfysize=2.5in \rotate[r]{\epsffile{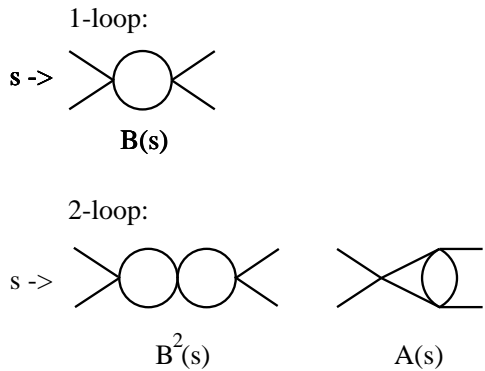}}
}
\vspace{0.2in}
\caption{The dominant $s$-channel high-energy topologies up to two
loops. The $t-$ and $u$-channel can be obtained using crossing relations.}
\label{feyn2lp}
\end{figure}

At high center-of-mass energy, $\sqrt{s}\gg M_H$, the three-point
interactions contributing to the amplitude $A$ are suppressed by
powers of $M_H^2/s$:
\begin{eqnarray}
A(W_L^+W_L^-\rightarrow Z_LZ_L) \;&=&\; 
-2\lambda \,-\, \frac{4\lambda^2v^2}{s-M_H^2} 
\,+\, {\rm O}(\lambda^2) \,+\, {\rm O}(\lambda^4v^4)\\
     &=&\; -2\lambda\left(1 + \frac{M_H^2}{s-M_H^2}\right) 
\,+\, {\rm O}(\lambda^2) \,+\, {\rm O}(\lambda^2 M_H^2)
\label{approx}\\
&\stackrel{s\gg M_H^2}{\rightarrow}&\; -2\lambda \,+\, {\rm O}(\lambda^2)
\,+\, {\rm O}\left(\frac{M_H^2}{s}\right).
\end{eqnarray}
This is also true when including higher order corrections, as can be
shown using dimensional arguments. Hence the entire high-energy
$s$-dependence originates from scattering graphs related to ${\cal
L}_{4pt}$, the part of the Higgs Lagrangian which is identical to a
massless $\phi^4$ theory with N=4.  The resulting high-energy
scattering graphs up to two loops are shown in Fig.~\ref{feyn2lp}.
Calculating these leading corrections up to two loops, the
renormalized OMS amplitude is \cite{DMR1,R}
\begin{eqnarray}
\label{wwzzcross}
\sigma(s)\, 
& = & \frac{1}{8\pi s} [\lambda(\mu)]^2\,
\Biggl[\,1\,+
\left( 24 \ln \frac{s}{\mu^2} -
       \, 42.65 \right) \,\frac{\lambda(\mu)}{16\pi^2}\,
 \\ 
&& \phantom{\frac{1}{8\pi s} [\lambda(\mu)]^2\, }
+ \left( 432 \ln^2 \frac{s}{\mu^2} - 1823.3\ln \frac{s}{\mu^2} 
   +\;2457.9\,\right) \frac{\lambda^2(\mu)}{(16\pi^2)^2}\,
+\; {\rm O}\left(\lambda^3(\mu)\right)\,\Biggr]\,,\nonumber
\end{eqnarray}
where the explicit $\mu$ dependence has been kept using the results in
\cite{NR}.  An anomalous dimension prefactor has been neglected since
it is close to unity for the values of $\sqrt s$ and $M_H$ considered
here.  Using the three-loop running coupling in connection with the
above two-loop result, we obtain the RG improved
next-to-next-to-leading-log (NNLL) cross section. The NNLL result
contains a complete summation of the leading logarithms (LL),
$\lambda^{k+1}\ln^k(s/\mu^2)$, the next-to-leading logarithms,
$\lambda^{k+2}\ln^k(s/\mu^2)$, and the NNLLs,
$\lambda^{k+3}\ln^k(s/\mu^2)$, $1\leq k\leq\infty$.  Just using the
one-loop cross section together with the two-loop running coupling
yields the NLL cross section, and the LL cross section corresponds to
the tree-level result with $\lambda(\mu)$ given by Eq.~(\ref{run1lp}).

The product $s\sigma$ depends only on the running
coupling  and the ratio $\mu/\sqrt s$, 
making the above cross section a
useful quantity in determining perturbative upper bounds on
$\lambda(\mu)$.
Taking $\mu=\sqrt{s}$, all RG logarithms are completely
resummed into $\lambda(\sqrt{s})$ and we obtain
\begin{eqnarray}
s\sigma\:\propto\:
\lambda^2(\sqrt{s})\left(1 - 42.65
\frac{\lambda(\sqrt{s})}{16\pi^2} 
+ 2457.9 \frac{\lambda^2(\sqrt{s})}{(16\pi^2)^2}\right)\,.
\label{cross}
\end{eqnarray}
It is striking that the coefficients appearing in the cross section
are much larger than the corresponding coefficients of the decay
widths, Eqs.~(\ref{hwwwidth}) and (\ref{hffwidth}).  Peculiarly, a
value $\lambda(\sqrt{s})>2.7$ leads to a {\it negative} NLL cross
section, indicating the complete failure of perturbation theory for
such a coupling.\footnote{ The value $\lambda(\sqrt{s})=2.7$
corresponds to, for example, $M_H=500$ GeV and $\sqrt{s}=1.2$ TeV.}
Using the various criteria listed earlier, the upper perturbative
bound is about $\lambda(\sqrt{s})\approx 2.2$ \cite{DJL,DMR2,R,NR}.This
upper bound on a perturbative Higgs running coupling represents a
strong restriction on the Higgs mass, stronger than the bound obtained
in the case of the decay widths, $\lambda(M_H)\approx 4$.  Choosing,
for example, $\sqrt{s}$ to be about 2 TeV, the Higgs mass has to
be less than about 400 GeV to allow for a perturbative calculation of
the cross section.  Recent results \cite{RW}, however, show that
this bound can be relaxed; see the following section.

\subsection{Improving perturbation theory}

\begin{figure}[tb]
\vspace*{13pt}
\centerline{
\epsfysize=2.5in \rotate[l]{\epsffile{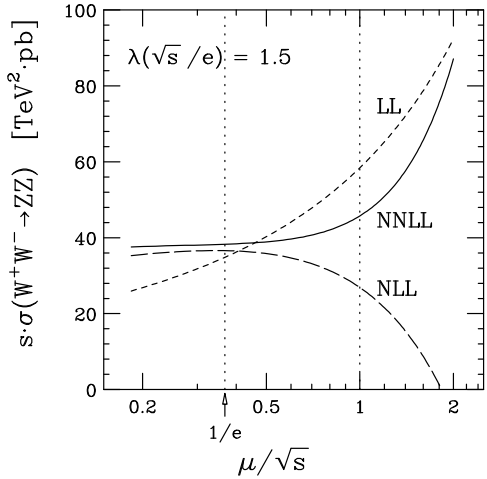}}
}
\vspace{0.2in}
\caption{The scale dependence of the high-energy cross section
$W_L^+W_L^-\rightarrow Z_LZ_L$. The running coupling is fixed to be
$\lambda(\protect\sqrt{s}/e)=1.5$, corresponding to 
$\lambda(\protect\sqrt{s})=1.9$.}
\label{figsigmu}
\end{figure}

The perturbative behaviour of the scattering cross sections can be
improved by performing a partial summation of higher-order
contributions \cite{RW}.  Realizing that chains of bubble diagrams,
$B(s)$, and diagrams with bubble substructures (see Fig.~\ref{feyn2lp}
for examples up to two loops) are the dominant contribution to the
$s$-dependent part of the scattering amplitude, a class of
nonlogarithmic terms accompanying the logarithmic $s$-dependent terms
can be partially summed.  This summation can be formulated in terms of a
modified scale entering the running coupling.  Instead of using the
standard choice $\mu=\sqrt{s}$, the improved choice is \cite{RW}
\begin{equation}
\mu=\frac{\sqrt s}{e}\approx\frac{\sqrt s}{2.7}\,.
\end{equation}
The vast improvement due to this choice can be seen when plotting the
scale dependence of the cross section. In Fig.~\ref{figsigmu} the
quantity $s\sigma$ is given as a function of $\mu/\sqrt{s}$, fixing
the running coupling at $\mu=\sqrt{s}/e$ to be 1.5 (which corresponds
to $\lambda(\sqrt{s})=1.9$).  
Taking $\mu=\sqrt{s}$, the cross section receives large radiative
corrections and has a strong scale dependence. In particular, the
scale dependence remains large even when performing a NNLL
resummation.  In contrast, the improved scale $\mu=\sqrt{s}/e$ leads
to much smaller corrections, and the scale dependence almost vanishes
when going beyond the LL approximation.  Of course, increasing the
value of the running coupling, perturbation theory also breaks down
for the summed cross section.  This happens for $\lambda(\sqrt{s}/e)$
of about 4.0, where perturbative unitarity is violated \cite{RW}.
Such a running couplings corresponds to, for example, $M_H\approx 700$
GeV and $\sqrt{s}$ close to 2 TeV.

\begin{figure}[tb]
\vspace*{13pt}
\centerline{
\epsfysize=2.5in \rotate[l]{\epsffile{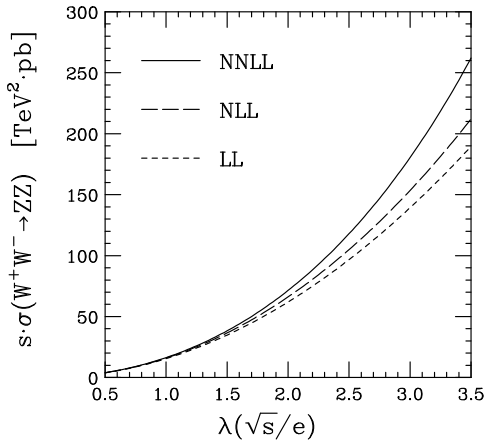}}
}
\vspace{0.2in}
\caption{
  The scaled cross section of $W^+W^-\rightarrow ZZ$ for
  $0.5<\lambda<3.5$ in the high-energy approximation, fixing
  $\mu=\protect\sqrt{s}/e$. Gauge and Yukawa coupling contributions are
  neglected.  }
\label{figsigall}
\end{figure}

Experimental measurement of the Higgs quartic coupling (for example, the
measurement of the $O(\lambda)$ contribution to high-energy
$W^+W^-$ scattering) and verifying its theoretical prediction is
challenging.  For small values of $\lambda$, the
electroweak $O(g^2)$ terms are dominant.  Larger values of $\lambda$
increase the $O(\lambda)$ contributions of the cross sections 
(see Fig.~\ref{figsigall}), but accumulation of a sufficient amount of data 
and elimination of background is a difficult task.

\section{Discussion}
\label{disc}

The perturbative upper bounds on the Higgs (running) coupling indicate
that perturbation theory is reliable for Higgs masses up to 700 GeV if
the process considered has a center-of-mass energy of less than 2 TeV.
Going to higher cms energy, the value of $M_H$ has to be reduced to
satisfy $\lambda(\sqrt{s}/e)<4.0$.  Of course, these bounds are
process dependent.  However, all $2\rightarrow2$ high-energy
scattering processes involving $W_L^+$, $W_L^-$, $Z_L$ and $H$
particles are related by the underlying SO(4) symmetry of ${\cal
L}_{4pt}$ and therefore have similar amplitudes.  In particular the
unitarity argument, first used by \cite{lee,LQT}, involves the different
channels and supports the above bound.

The perturbative upper bounds can be compared with results obtained
from lattice calculations \cite{LW2,LW1,GKNZ}.  Such results depend
somewhat on the lattice action and regularization used.
Interestingly, they obtain similar bounds on the Higgs mass, excluding
the existence of a heavier standard model Higgs boson.  Together with
the results reviewed here, this excludes the possibility of a strongly
interacting Higgs boson \cite{RW}.

\vspace{1cm}

\bigskip
\centerline{\bf ACKNOWLEDGEMENTS}
\smallskip\noindent
The author thanks the organizers of the {\it XXXVI Cracow School of 
Theoretical Physics} for the invitation to this informative and well-organized
school. The pleasant atmosphere as well as the picturesque surroundings
provided a great setting for lectures and discussions.


\newpage

\begin{thebibliography}{99}

\bibitem{goldstone}
J. Goldstone, Nuovo Cimento {\bf 54}, 154 (1961); T.-P. Cheng and 
L.-F. Li, {\em Gauge Theory of Elementary Particle Physics}
(Oxford University Press, Oxford, Englend, 1984), Chap.~5.

\bibitem{lee}
B.W. Lee, C. Quigg, and H.B. Thacker, Phys.\ Rev.\ Lett.\ {\bf 38},
883 (1977); Phys.\ Rev.\ D {\bf 16}, 1519 (1977).

\bibitem{eqt1}
M.S. Chanowitz and M.K. Gaillard, Nucl.\ Phys. {\bf B261}, 379 (1985).

\bibitem{eqt2}
J. Cornwall, D. Levin, and G. Tiktopoulos, Phys.\ Rev.\ D {\bf 10}, 1145 
(1974), Appendix A; C.E. Vayonakis, Lett.\ Nuovo Cimento
{\bf 17}, 383 (1976); G. Gounaris, R. K\"{o}geler, and H. Neufeld,
Phys.\ Rev.\ D {\bf 34}, 3257 (1986); Y. Yao and C. Yuan, {\em
ibid}. {\bf 38}, 2237 (1988).
%
\bibitem{BS}
J. Bagger and C. Schmidt, Phys.\ Rev.\ D {\bf 41}, 264 (1990);
A. Aoki, Kyoto University Report No. RIFP-705, 1987 (unpublished):
in {\em Physics at TeV Scale}, Proceedings of the Meeting, Tsukuba,
Japan, 1988, edited by K. Hidaka and K. Hikasa (KEK, Tsukuba,
1988); H. Veltman, Phys.\ Rev.\ D {\bf 41}, 2294 (1990).

\bibitem{LW2} M.~L\"uscher and P.~Weisz, \NPB 318 705 1989 .

\bibitem{NR} U.~Nierste and K.~Riesselmann, \PRD 53 6638 1996 .

\bibitem{sher} M. Sher, Phys.\ Rep.\ {\bf 179}, 273 (1989).

\bibitem{LW1} M.~L\"uscher and P.~Weisz, \PLB 212 472 1988 .

\bibitem{GKNZ} M.~G\"ockeler, H.~Kastrup, T.~Neuhaus, and F.~Zimmermann, 
\NPB 404 517 1993 .

\bibitem{hww1} A.~Ghinculov, \NPB 455 21 1995 .

\bibitem{hww2} A. Frink, B.A. Kniehl, D. Kreimer, and K. Riesselmann,
Report Nos.\ TUM--HEP--247/96 
and hep--ph/9606310 (May 1996); to appear in Phys.\ Rev. D.

\bibitem{hff1} L.~Durand, B.A.~Kniehl, and K.~Riesselmann, Phys.\ Rev.\ Lett.\ 
  {\bf72}, 2534 (1994); {\bf74}, 1699(E) (1995); Phys.\ Rev.\ D {\bf51}, 5007
  (1995). 
  
\bibitem{hff2} A.~Ghinculov, \PLB 337 137 1994 ; (E) {\bf 346}, 426 (1995).

\bibitem{DMR1} P.~Maher, L.~Durand, and K.~Riesselmann, \PRD 48 1061
1993 ; (E) {\bf 52}, 553 (1995).

\bibitem{R} K.~Riesselmann, \PRD 53 6226 1996 .


\bibitem{DJL} L.~Durand, J.~Johnson, and J.~Lopez, \PRL 64  1215 1990 ;
\PRD 45 3112 1992 .


\bibitem{DMR2} L.~Durand, P.~Maher, and K.~Riesselmann, 
\PRD 48 1084 1993 .

\bibitem{RW}      
   K. Riesselmann and S. Willenbrock,
   Report No. TUM--HEP--236/96 (August 1996) and hep/ph9608280.


\bibitem{LQT} D.~Dicus and V.~Mathur, \PRD 7 3111 1973 .

\end{thebibliography}
\end{document}